\documentclass[11pt]{article}

\usepackage[utf8]{inputenc}
\usepackage[english]{babel}
\usepackage{amsmath}
\usepackage{color,soul}
\usepackage{graphicx}
\usepackage{lipsum}

\usepackage{subfiles}
\usepackage{subcaption}
 
\usepackage{blindtext}
\usepackage{etoolbox}

\usepackage{booktabs}
\usepackage{hyperref}
\hypersetup{
    colorlinks,
    citecolor=red,
    filecolor=black,
    linkcolor=black,
    urlcolor=black
}
\usepackage{float}

\begin{document}

\newcommand{\mainfile}{}
 
\begin{titlepage}

\centering
{\scshape\LARGE Technical Report \par}
\vfill
{\huge\bfseries {Towards Learned Predictability of Storage Systems} \par}
\vspace{2cm}
{\scshape\Large {Chenyuan Wu} \par}
\vspace{1cm}
{\scshape\Large Department of Computer and Information Science\par}
{\scshape\Large University of Pennsylvania\par}
\vfill
{\large \today \par}
\end{titlepage}


\begin{abstract}

With the rapid development of cloud computing and big data technologies, storage systems have become a fundamental building block of datacenters, incorporating hardware innovations such as flash solid state drives and non-volatile memories, as well as software infrastructures such as RAID and distributed file systems. Despite the growing popularity and interests in storage, designing and implementing reliable storage systems remains challenging, due to their performance instability and prevailing hardware failures.

Proactive prediction greatly strengthens the reliability of storage systems. There are two dimensions of prediction: performance and failure. Ideally, through detecting in advance the slow I/O requests, and predicting device failures before they really happen, we can build storage systems with especially low tail latency and high availability. While its importance is well recognized, such proactive prediction in storage systems, on the other hand, is particularly difficult. To move towards predictability of storage systems, various mechanisms and field studies have been proposed in the past few years. In this report, we present a survey of these mechanisms and field studies, focusing on machine learning based "black-box" approaches. Based on three representative research works, we discuss where and how machine learning should be applied in this field. The strengths and limitations of each research work are also evaluated in detail.

\end{abstract}
	
	\ifcsdef{mainfile}{}{\bibliography{../references/main}}

\pagebreak

\tableofcontents

\pagebreak

\section{Introduction}
\label{sec:introduction}

    With the rapid development of cloud computing and big data technologies, storage systems have become a fundamental building block of datacenters, incorporating hardware innovations such as flash Solid State Drives (SSDs) and Non-Volatile Memories (NVMs), as well as software infrastructures such as RAID and distributed file systems. Despite of the growing popularity and interests in storage, designing and implementing \textit{reliable} storage systems are still challenging. One major reason is performance instability: RAIDs suffer 1.5\% (for HDDs) and 2.2\% (for SSDs) of RAID hours with at least one slow drive, which is 2x slower that its peer drives\cite{hao2016tail}. This indicates stable latencies at 99th percentile is hard to achieve in current RAID deployments, let alone the stricter application level SLOs. The other reason is prevailing hardware failures: the annualized failure rate is reported to be 1.36\% for HDDs\cite{lu2020making}, and 3.92\% for SSDs\cite{han2021depth,xugeneral}. Such high level of failure rates along with the massive scale of hardware fleets require careful device level health monitoring.
    
    Proactive prediction provides a new angle to achieve reliable storage systems in two dimensions: performance and failure. By detecting in advance the I/O requests whose performance expectations cannot be fulfilled, we can eventually cut the tail latencies and thereby guarantee SLOs; by predicting HDDs/SSDs failures before they really happen, we can achieve high availability even at the device level. It is well recognized that proactive prediction is integral for developing reliable storage systems in commercial datacenters, and it can act as a plug-in complement for existing approaches that aim to achieve reliable storage in datacenters(e.g., hedged requests, replication and RAIDs). While its importance is recognized, predictability of storage systems is particularly difficult, mainly due to the heterogeneity of drives in deployment, and the complex internal idiosyncrasies of modern storage devices.
    
    There are some popular approaches to mask such unpredictability, i.e. live with and embrace this unpredictability. A common practice for masking performance unpredictability is "hedged requests"\cite{dean2013tail}, which sends a duplicated I/O request to another node if it has been outstanding for more than the 95th percentile of expected latency. However, this approach increases the I/O tensity and results in poor resource utilization. A common practice for masking failure unpredictability is replication and erase coding, which reconstructs corrupted chunks using healthy data and parity chunks. However, faced with prevalent correlated failures\cite{han2021depth}, such redundancy schemes are shown to be insufficient. All these facts suggest that researchers should move \textit{towards the predictability of storage systems} instead of escaping from them.
    
    Over the past few years, researchers have devoted intensive effort in exploring proactive and accurate prediction of performance and failure in storage systems. Various techniques have been proposed for storage performance prediction, including "white-box" approaches that require detailed understanding of the device\cite{hao2017mittos}, as well as ML based "black-box" approaches\cite{hao2020linnos}. For storage failure prediction, researchers have proposed several ML based techniques for both SSDs\cite{xugeneral,mahdisoltani2017proactive} and HDDs\cite{lu2020making,han2020toward,mahdisoltani2017proactive}. In this paper, we are going to provide a survey of \textbf{machine learning based "black-box" approaches} for prediction tasks in storage systems.
    
    The rest of this paper is organized as follows. We start with problem statement in Section~\ref{sec:problemstatement}, by introducing the storage hierarchy in datacenters and defining the specific prediction tasks. Section~\ref{sec:taxonomy} provides a taxonomy of the mechanisms and field studies for prediction tasks in storage systems, according to the techniques adopted. In Section~\ref{sec:linnos}-\ref{sec:smarter}, we then present three representative systems/field studies that aim to achieve predictability in storage systems from different aspects. Finally, Section~\ref{sec:discussion} reviews all the mechanisms and field studies and discusses challenges.

    \ifcsdef{mainfile}{}{\bibliography{../references/main}}

\pagebreak

\section{Problem Statement}
\label{sec:problemstatement}
	
    To scope our survey paper, we begin first with an overview of the storage systems we are looking at, then define three specific prediction tasks that this paper focuses on.
    
    \subsection{Storage Hierarchy in Datacenters}
    \label{sec:2.1}
    \begin{figure}[H]
        \centering
        \includegraphics[scale=0.35]{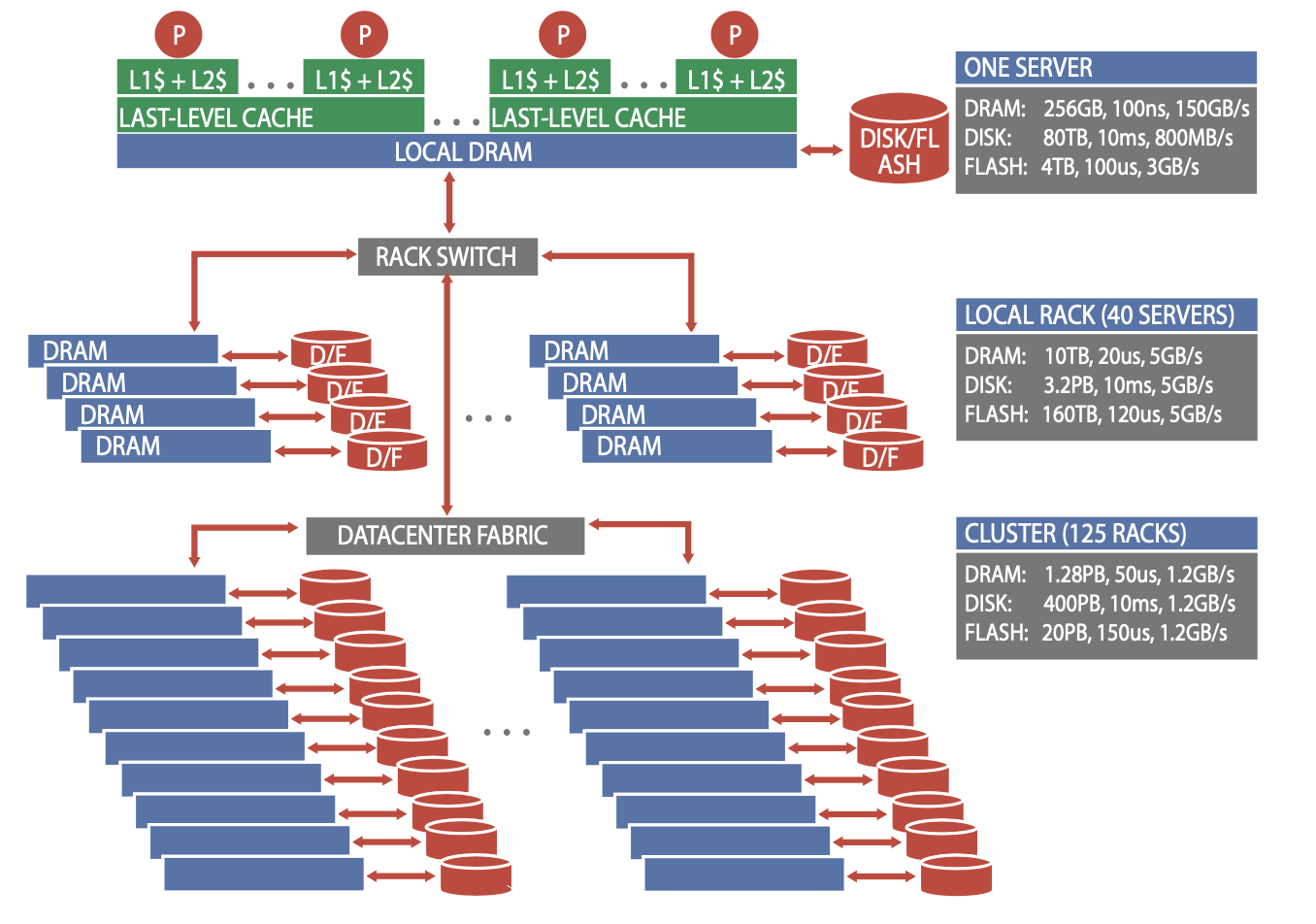}
        \caption{Storage Hierarchy of a Datacenter \cite{barroso2018datacenter}}
        \label{fig:hierarchy}
    \end{figure}
    
    Figure~\ref{fig:hierarchy} shows a programmer's view of storage hierarchy of a datacenter. The server consists of a number of processor sockets, each with a multicore CPU and its internal cache hierarchy, local DRAM, a number of directly attached disk drives or flash based solid state drives. The DRAM and disk/flash drives within the rack are accessible through the first-level rack switches (assuming some sort of remote procedure call API to them exists), and all resources in all racks are accessible via the cluster level switch. The capacity, latency, throughput of each level of storage is shown in the figure.
    
    To increase throughput and robustness at the local drive level, \textit{RAID} technologies are widely deployed, where multiple homogeneous drives are connected to the host server via a RAID controller. To further increase capacity and robustness, and hide the complexity of using multiple remote memories and remote disk/flash devices, \textit{distributed file systems} are proposed, e.g. NFS, GFS, Colossus, etc. Due to their almost unlimited capacity and high level of fault tolerance, distributed file systems are widely used as the storage backend for \textit{big data applications}. For instance, each worker node in Spark\cite{zaharia2012resilient,zaharia2010spark} (similar as the server at the highest level in Figure~\ref{fig:hierarchy}) is connected to a master node for coordination and the HDFS for storage (similar as the lower two levels in Figure~\ref{fig:hierarchy}), and data can be shuffled between multiple worker nodes. When materializing an RDD, Spark will first try to cache it within the distributed memory; if insufficient, some partitions will be stored or evicted to HDFS. Note that different prediction tasks may focus on different part of this hierarchy, which will be elaborated in the following subsection.
    
    \subsection{Task Definitions}
    \label{sec:2.2}
    
    \begin{figure}[H]
        \centering
        \includegraphics[scale=0.55]{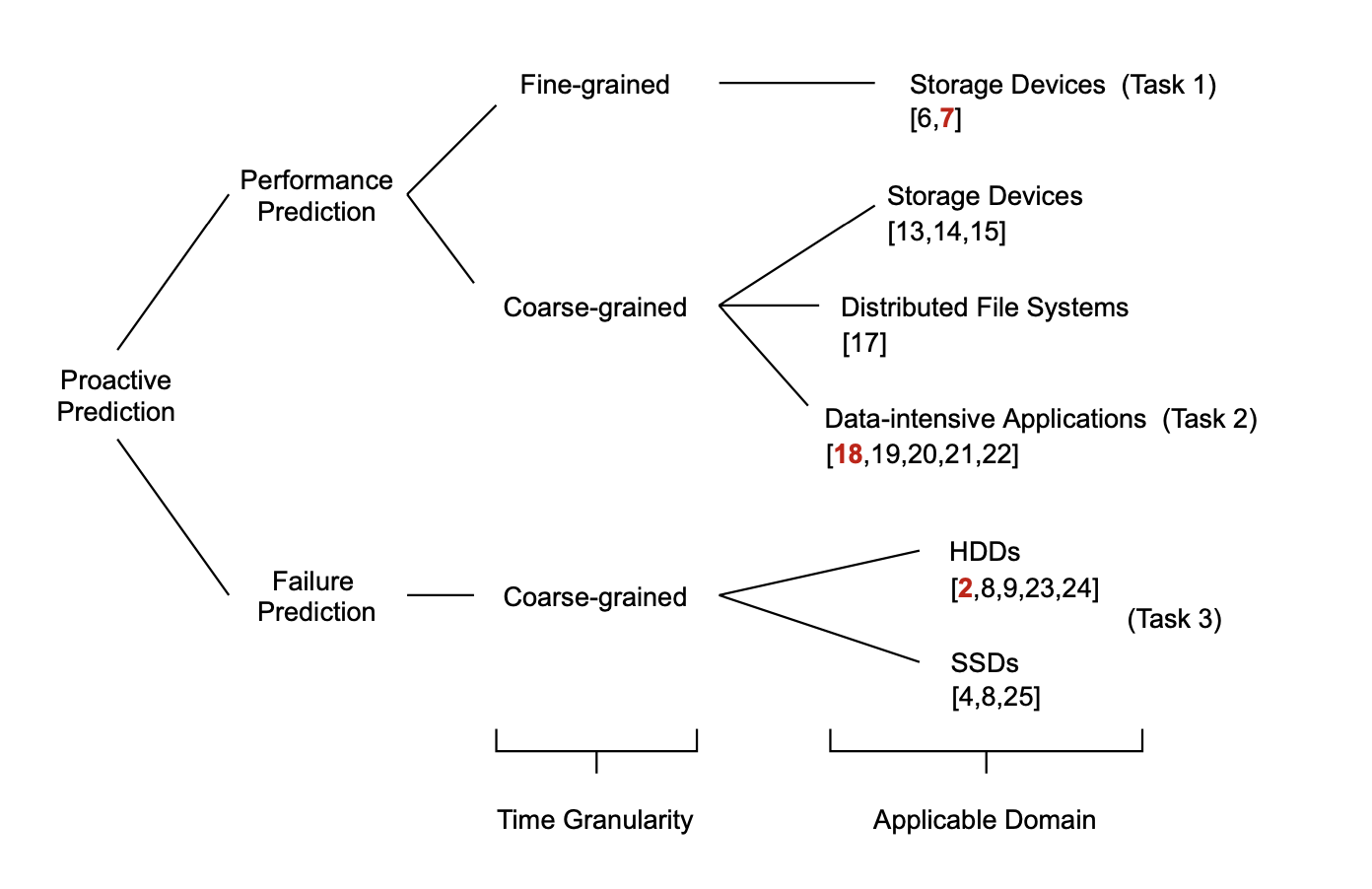}
        \caption{Overview of Different Prediction Tasks}
        \label{fig:task tree}
    \end{figure}
    
    Figure~\ref{fig:task tree} presents an overview of different proactive prediction tasks, each uniquely defined by its \textit{time granularity} and \textit{applicable domain}. By time granularity, we mean how often we need to make such prediction, which further imposes an upper bound on the inference time. More specifically, fine-grained prediction indicates the inference time should be sub-ms or even sub-us scale, while coarse-grained prediction indicates the inference time could be around a few seconds. By applicable domain, we mean which part in the storage hierarchy (as in Section~\ref{sec:2.1}) this task is focusing on. In total, there are six different tasks according to existing literature, as shown in Figure~\ref{fig:task tree}.
    
    Due to space limit, among these six different tasks, we choose four \textit{recently well studied} ones (captioned as Task 1-3 in Figure~\ref{fig:task tree}) for closer inspection. Note that we integrate the two failure prediction tasks (for HDD/SSD respectively) into Task 3 due to their similarity in usage and requirements. Coarse-grained performance prediction for storage device\cite{wang2004storage,yin2006empirical,varki2004issues} was well studied twenty years ago, as a helper of the solver in automated storage management tools like Hippodrome\cite{anderson2002hippodrome}. However, its importance has drastically faded with the emergence of cloud computing and various distributed file systems that are well designed for different workloads. On the other hand, coarse-grained performance prediction for distributed file systems\cite{hsu2016inside} is unluckily receiving little research attention. As a consequence, we didn't include them in our closer inspection.
    
    For each of the chosen tasks, we first describe at a high level the usage of the prediction as well as requirements on the input and output, all of which are imposed by the time granularity and applicable domain of that task. Later, we present a case study for each chosen tasks in Section~\ref{sec:linnos}-\ref{sec:smarter}.
    
    \subsubsection{Task 1: Fine-grained Performance Prediction for Storage Devices}
    \label{sec:task 1}
    In this task, since we make prediction and take corresponding action for \textit{each incoming I/O on the target storage device}, the inference time should be negligible comparing to the original I/O latency: a $<\%3$ overhead per I/O is desirable. Thus, for hard disk drives, the inference time should be sub-1ms, for solid state drives, the inference time should be sub-10us.
    
    \begin{itemize}
        \item \textbf{Usage:} On each device, we want to reject the incoming I/O requests whose SLOs cannot be met, and perform failover to another less busy device within the same RAID array or on another server (in the case of NoSQL systems). The failover has little overhead (only $\sim$15us within the same RAID), comparing to the time spent on waiting otherwise. By performing this quick detection and failover, we cut the "tail" in datacenters.
        \item \textbf{Input:}  In order to meet the requirements on inference time, the input features to the performance model should be minimized to only include those that really matters.
        \item \textbf{Output:} The output can be either a binary label (e.g., whether the SLO can be met), or the precise value of the metric (e.g., I/O latency) which can then be compared with the SLO provided by the application.
    \end{itemize}
    
    Examples of this task include\cite{hao2020linnos,hao2017mittos}. Section~\ref{sec:linnos} is a case study of this task.
    
    \subsubsection{Task 2: Coarse-grained Performance Prediction for Data-intensive Applications}
    \label{sec:task 2}
    In this task, we make prediction for each incoming job, before they are submitted or scheduled. Since a job in big data frameworks (e.g., MapReduce, Spark) always lasts for tens of seconds, we don't have a strict upper bound on the inference time. It would be best if the inference time can be controlled to sub-second scale.
    
    \begin{itemize}
    \item \textbf{Usage:} The prediction should be made at the time of planning and scheduling a job. For datacenter administrators or resource management software in datacenters, the prediction result helps to determine where to launch this new, say Spark, instance so that the SLO won't be violated; for cloud service users, it helps them choose the cheapest VM configuration given a performance goal, or choose the most performant one given a budget. 
    \item \textbf{Input:} Since the prediction is made before the job starts running, static input features are required, i.e. they should be known prior to running. This suggests that no runtime counters or runtime measurements (e.g., CPU utilization, average I/O latency) can serve as the input.
    \item \textbf{Output:} The output should be exact metric such as JCT, query latency, etc.
    \end{itemize}
    
    Although coarse-grained prediction sounds easier, at least similar, to fine-grained prediction in Task 1, more challenges arise for performance prediction at the level of data-intensive applications: 1) The datacenter network is involved in both storage backend (distributed file systems) and the data shuffling between compute nodes, as illustrated in section~\ref{sec:2.1}, which is a common source of contention and thus the source of unpredictability; 2) The application itself may contain randomizations and intricate system optimizations, which are also the source of unpredictability. 
    
    Examples of this task include\cite{fu2021use,ousterhout2017monotasks,venkataraman2016ernest,yadwadkar2014wrangler,verma2011aria}. Section~\ref{sec:mlblackbox} is a case study of this task. 
    
    \subsubsection{Task 3: Coarse-grained Failure Prediction for HDDs and SSDs}
    \label{sec:task 3}
    In this task, we explore the second dimension of proactive prediction: failure. We make failure prediction for every already deployed HDD and SSD on a daily basis. Therefore, there is no strict requirement on the inference time. However, unlike previous tasks, here the prediction horizon is much larger, e.g., 10-20 days should suffice. This means we are predicting whether this drive would fail within the next 10-20 days.
    
    \begin{itemize}
    \item \textbf{Usage:} After predicting whether a certain drive is going to fail in the next few days, the datacenter administrators can further manually examine or replace all suspicious drives. By fixing these suspicious drives, we can prevent correlated failures (i.e., failures within the same spatial domain and the same time period) from happening, for which current redundancy schemes are insufficient. 
    \item \textbf{Input:} SMART attributes together with other informative features should be used as input. Here we are able to use fancier features and summary statistics as input.
    \item \textbf{Output:} The output should be a binary label.
    \end{itemize}
    
    Although failure and performance looks unrelated to each other, they are never strictly decoupled in the systems filed, since failures often adversely affect performance\cite{hao2016tail}, and performance is a strong indicator for failures as we will show later. Examples of this task include \cite{lu2020making,xugeneral,han2020toward,mahdisoltani2017proactive,xiao2018disk,xu2018improving,alter2019ssd}. Section~\ref{sec:smarter} is a case study of this task. 
    
    \ifcsdef{mainfile}{}{\bibliography{../references/main}}

\pagebreak

\section{Taxonomy of Techniques}
\label{sec:taxonomy}
	
\begin{figure}[H]
    \centering
    \includegraphics[scale=0.35]{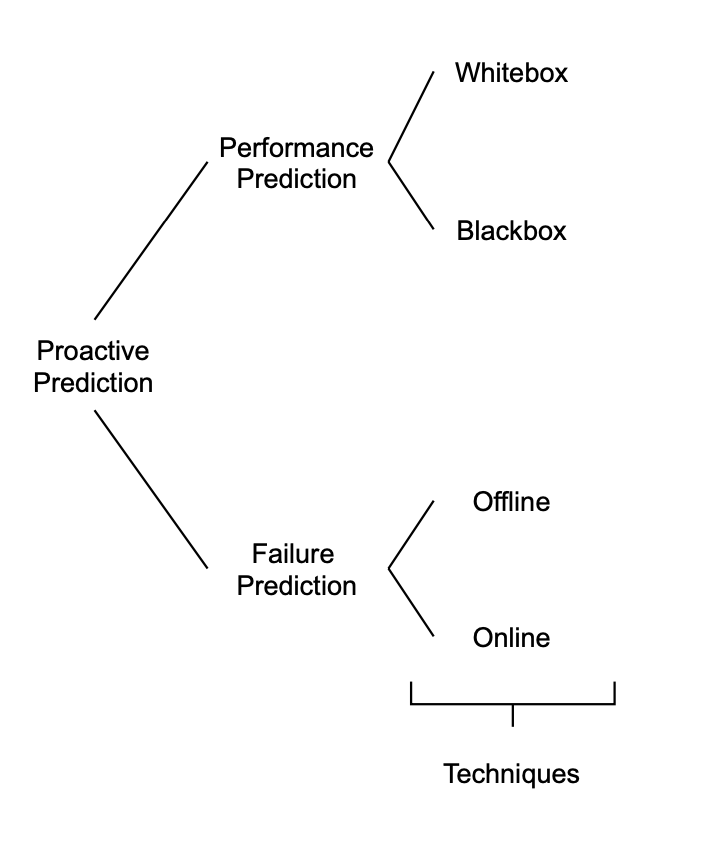}
    \caption{Classification of Prediction Techniques}
    \label{fig:technique tree}
    \end{figure}

    Having presented the definitions of proactive prediction, we further present a taxonomy of different prediction \textit{techniques} proposed (shown in Figure~\ref{fig:technique tree}). Note that the techniques surveyed in this section are orthogonal to task definitions in Section~\ref{sec:problemstatement}, i.e., for a certain task, any technique applies as long as it meets the requirements. For instance, either whitebox\cite{hao2017mittos} or blackbox\cite{hao2020linnos} approach can be applied to Task 1.
    
    \subsection{Performance Prediction}
    \label{sec:3.1}
    Performance prediction techniques can be taxonomized as whitebox\cite{hao2017mittos, varki2004issues, ousterhout2017monotasks, verma2011aria} or blackbox\cite{hao2020linnos,wang2004storage,yin2006empirical, fu2021use,venkataraman2016ernest,hsu2016inside,yadwadkar2014wrangler}, according to how they view the underlying device or system, which is further discussed below.
    
    \begin{itemize}
        \item \textbf{Whitebox:} Whitebox approaches assume full knowledge of the underlying system internals. For instance, MittOS\cite{hao2017mittos} built its prediction model based on its understanding of contention and queueing discipline of the underlying resource (e.g., disk spindles vs. SSD channels, FIFO vs. priority). Some whitebox approaches may even re-architect the system in order to build a better prediction model\cite{ousterhout2017monotasks}.
        
        \item \textbf{Blackbox:} Blackbox approaches have no information about the internal components or algorithms of the underlying system. Thus, most blackbox approaches utilize machine learning to learn the systems behavior from examples. 
        
        On the one hand, blackbox approaches are more generalizable and scalable than whitebox ones. Often, the device internals are proprietary and not exposed to researchers. Further, it's almost infeasible to deploy whitebox approaches in production deployment, since numerous vendors - each with various device types - are involved, where each type of device need expertise's inspection in order to build the whitebox model. On the other hand, blackbox approaches are often fundamentally limited, resulting in poor prediction accuracy even under exceptionally simplified settings, as is later shown in Section~\ref{sec:mlblackbox}. 
    \end{itemize}
    
    \subsection{Failure Prediction}
    \label{sec:3.2}
    Most failure prediction techniques are blackbox approaches, and they can be further taxonomized as offline\cite{xugeneral,mahdisoltani2017proactive,alter2019ssd,lu2020making,xu2018improving} or online\cite{han2020toward,xiao2018disk}, according to how the performance model is trained or built, which is further discussed below.
    
    \begin{itemize}
        \item \textbf{Offline:} Offline scheme means that all training data must be available before training the prediction model. In other words, the model won't upgrade once it is deployed in production.
        \item \textbf{Online:} Online scheme means that the prediction model is updated incrementally in real time (i.e., after being deployed), upon receiving each newly generated healthy or failure sample. Some work utilizes online random forest algorithm\cite{xiao2018disk}, while other work formulates a stream mining problem so as to combat concept drift\cite{han2020toward}.

        Online approaches are generally more preferred than offline approaches, since disk logs are continuously generated, in which the statistic patterns may vary over time. That being said, offline schemes may suffer in long-term use, because the testing data is going to come from a different distribution from training data.
    \end{itemize}
    
    \ifcsdef{mainfile}{}{\bibliography{../references/main}}

\pagebreak

\section{LinnOS: Predictability on Unpredictable Flash Storage with a Light Neural Network}
\label{sec:linnos}
	
    LinnOS is an operating system that leverages a light neural network for inferring SSD performance at a very fine--per I/O--granularity and helps parallel storage applications achieve performance predictability. LinnOS supports blackbox devices and real production traces without any extra input from users, while outperforming industrial mechanisms and other approaches.
    
    \subsection{Overview}
    \label{sec:4.1}
    \noindent {\textbf{Usage Scenario:} LinnOS is beneficial for parallel, redundant storage such as flash arrays (cluster-based or RAID) that maintain multiple replicas of the same block. When a storage application performs an I/O via OS system calls, it can add a one-bit flag, hinting to LinnOS that the I/O is latency critical, and thus triggering LinnOS to infer the I/O latency. Before submitting the I/O to the underlying SSD, LinnOS inputs the I/O information to the neural network that it has trained, which will make a binary inference: fast or slow. If the output is "fast", LinnOS submits the I/O down to the device. Otherwise, if it is "slow", LinnOS revokes the I/O (not entered to the device queue) and returns a "slow" error code. Upon receiving the error code, the storage application can failover the same I/O to another replica. In the worst case, where the application must failover to the last replica, this last retry will not be tagged as latency critical so that the I/O will eventually complete and not be revoked.} \\
    
\begin{figure}[H]
    \centering
    \includegraphics[scale=0.45]{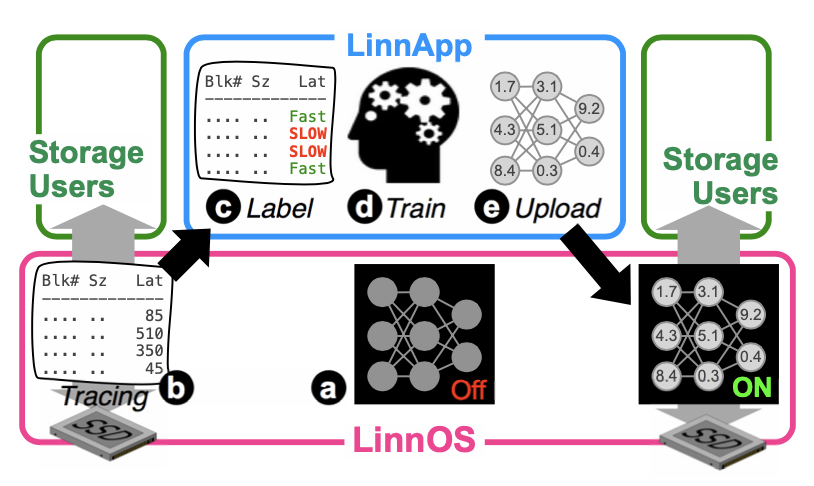}
    \caption{LinnOS Architecture}
    \label{fig:linnos architecture}
    \end{figure}

    \noindent {\textbf{Architectural Overview:} Figure~\ref{fig:linnos architecture} shows LinnOS's overall architecture. At the center of LinnOS is the speedy neural network that lies in the kernel space and infers the speed of every incoming I/O individually. To train the model, LinnOS collects and uses the current live workload that the SSD is serving. The collected trace is then supplied to LinnApp, a supporting user-level application. LinnApp runs an algorithm that automatically labels the traced I/Os with either "fast" or "slow", and proceeds with the training phase. The training phase generates the weights for the neurons in the model that will be uploaded to LinnOS. The model is then activated, and LinnOS is ready to inferences and revoke "slow" I/Os. }

    \subsubsection{Data Collection}
    \label{sec:4.1.1}
    I/O systems inherently can collect a large amount of data with low-overhead tools, which is necessary for training. LinnApp collects traces for every load-SSD pair to model. For example, for inferring a production workload performance on a particular SSD in deployment, an online trace will be collected. For every I/O, five raw fields are collected: the submission time, block offset, block size, read/write, and the I/O completion time. In this phase, raw fields are also converted to the input feature format.

    The main challenge here is to decide how long the trace should be. LinnOS takes a simple approach where it uses a busy-hour trace (e.g., midday). It is verified in the evaluation that for production workloads, as busy-hour trace well represents the others, i.e., the inflection point does not deviate significantly.
    
    \subsubsection{Labeling}
    \label{sec:4.1.2}
    \begin{figure}[H]
    \centering
    \includegraphics[scale=0.4]{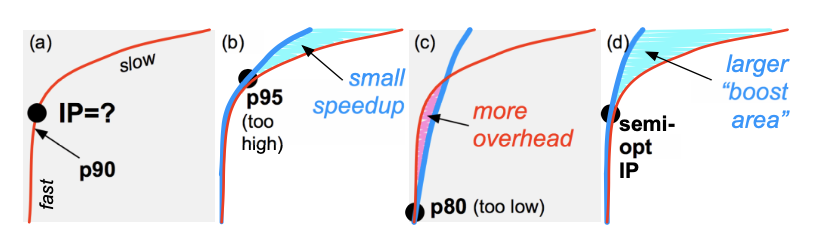}
    \caption{Inflection Point (figure format: latency CDF)}
    \label{fig:inflection point}
    \end{figure}   
    
    As LinnApp employs a supervised learning approach and does not require extra input from users, the training data must be automatically labeled. However, if every I/O is labeled with the actual us-level latency, the device behavior will be too hard to learn. Luckily, what users worry about is the tail behavior, not the precise latency: SSD latencies often form a high alpha Pareto distribution (as shown in Figure~\ref{fig:inflection point}a), where 90\% of the time, the latency is highly stable, but in the other 10\% of the time, it starts forming a long tail. Thus, LinnApp aims to automatically find the best inflection point (IP), and label each I/O as "fast" or "slow" accordingly. This binary labeling simplifies the problem, and therefore improves accuracy.

    Due to the hardware heterogeneity and unbalanced user load across devices, the inflection point differs for each load-device pair. For this reason, LinnApp collects per-device traces, finds per-device inflection point, and trains the model for every load-device pair in the array. 
    
    Assume during data collection, $t$ workload traces ($T_1$ to $T_t$) running on $d$ devices ($D_1$ to $D_d$) are collected, where $t==d$. The following best-effort algorithm is adopted to find a semi-optimum inflection point for each $T_i-D_i$ pair: (1) For each $T_i-D_i$ pair, we pick a starting IP value where the slope of the CDF is one. (2) For the current device $D_1$, we run a simulation of one million I/Os, where each I/O request $r_i$ takes a random latency value from $T_1$'s real latency distribution. We then simulate LinnOS admission control and decide the new latency: if $r_i$'s latency is smaller than the current IP, the new latency is the same; if not, $r_i$ will be revoked and failover to another randomly selected node (e.g., $D_4$) where a random latency is picked from its trace, and the admission control is repeated. (3) These new, optimized latencies of $r_i$ form the new CDF. Using the original and new CDFs, we can calculate the area difference (the shaded "boost area" in Figure~\ref{fig:inflection point}d), which represents the latency gain if use this IP value. (4) Still for $D_1$, we repeat all the steps above by moving +/-0.1 percentile within the +/-10 percentile ranges from the initial IP value. Then, we pick the IP value that gives us the largest boost area. (5) We repeat all the steps for other devices. 
    
    \subsubsection{Light Neural Network Model}
    \label{4.1.3}
    \begin{figure}[H]
    \centering
    \includegraphics[scale=0.4]{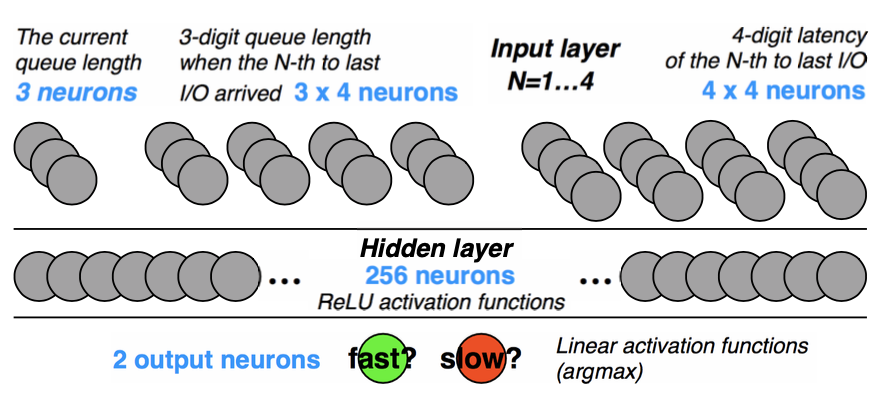}
    \caption{Light Neural Network}
    \label{fig:lnn}
    \end{figure}  
    
    Figure~\ref{fig:lnn} depicts the light neural network used, along with the input features and format. To infer the speed of every I/O, the model takes three inputs: (1) the number of pending I/Os when an incoming I/O arrives, (2) the latency of the $R$ most recently completed I/Os, where $R$ is set as 4, (3) the number of pending I/Os at the time when each of the $R$ completed I/Os arrived. Here $R$ is tradeoff between accuracy and inference time. These input features are minimized (i.e., not including block offset, write/read flag, and a long history, etc.) so as to reduce the inference time.

    To balance the model size and learning difficulty, the number of pending I/Os are formatted into three decimal digits. For example, the format for 15 pending I/Os is three integers {0,1,5}. Similarly, the us latencies of the recent completed I/Os are formatted into four digits.

    The final model is a fully connected neural network with only three layers. The input layer is supplied with the 31 features described above. The hidden layer consists of 256 regular neurons and uses RELU activation functions. The output layer has two neurons with linear activation function. This light design is crucial for the live, fine-grained prediction task.
    
    \subsubsection{Improving Accuracy}
    \label{sec:4.1.4}
    To further improve the model accuracy, \textit{false submit reduction}, \textit{model recalibration}, \textit{inaccuracy masking} is performed. Since the wrong inference penalty (in terms of its effect for cutting tail latency) is small for false revokes but high for false submits, biased training is used for reducing false submits. Model recalibration is used to adapt to significant workload changes, where re-tracing and re-computation of inflection point is done periodically every few hours. Inaccuracy masking combines LinnOS with hedging, where it uses the false submit rate as an indicator for the hedging percentile value.
    
    \subsubsection{Improving Inference Time}
    In addition to the 3-layer design that is fundamental for reducing inference time, LinnOS makes further optimizations. First, it adopts DNN quantization, where the trained floating-point weights are converted to integers with precision of three decimal points. Second, it can opportunistically use co-processors to reduce the average inference time with 2-threaded optimized matrix multiplication using one additional CPU core.
    
    \subsection{Evaluation}
    LinnOS is a case study for Task 1. According to the taxonomy presented in Section~\ref{sec:taxonomy}, it adopts blackbox techniques for performance prediction. We summarize the strengths and limitations of LinnOS as follows:\\
    
    \noindent \textbf{Strengths:}
    \begin{itemize}
        \item LinnOS is the first operating system that successfully infers I/O speed in a fast, accurate, fine-grained, and general fashion. It helps storage applications cut high percentile latency, thus achieving predictable performance on flash arrays.
        \item The auto-labeling algorithm exempts users from supplying an SLO value such as a deadline. Properly setting the SLO is not easy for users.
        \item With the help of recalibration, LinnOS can adapt to the workload shift that happens slowly and continuously in modern datacenters.
    \end{itemize}
    
    \noindent \textbf{Limitations:}
    \begin{itemize}
        \item In order to combat heterogeneity, LinnApp collects per-device traces and trains the model for every load-device pair in the array. However, this limits the scalability of LinnOS: a single host cannot be connected to a large RAID array, because a single host cannot hold the large training burden of multiple models, and the CPU overhead the models impose even during inference time.
        \item Though LinnOS inference overhead is less noticeable compared with the access latency of current SSDs, it could become problematic as SSDs march to 10us latency range. Also, the consumption of computation resources can increase substantially as the IOPS grow.
    \end{itemize}

    \ifcsdef{mainfile}{}{\bibliography{../references/main}}

\pagebreak

\section{On the Use of ML for Blackbox System Performance Prediction}
\label{sec:mlblackbox}
	
    Unlike LinnOS that focuses on a particular use case of ML-based performance prediction, Fu et.al. conducted this field study (referred to as MLSys), attempting to answer a broader question: does ML make prediction simple and general? MLSys develops a methodology for systematically diagnosing whether, when and why ML does (not) work for performance prediction, and identify steps to improve predictability.
    
    \subsection{Overview}
    \label{sec:5.1}
    \noindent {\textbf{Methodology:} In what follows, we first introduce the metrics and parameters, followed by two tests and two predictors that are used throughout MLSys.
    
    \textit{Metrics:} For generality and remaining agnostic to the specifics of a use-case, prediction quality is measured using the root mean square relative error (rMSRE).
    
    \textit{Parameters:} MLSys considers the following three classes of parameters that impact an application's performance: application-level input that the application acts on, including both the size of these inputs and (when noted) the actual values of these inputs; application-level configuration that is exposed to users to tune its behavior, e.g., the degree of parallelism; infrastructure that captures the resources on which the application runs, e.g., CPU speed, memory size, etc. These are all static parameters that can be known prior to running the application.
    
    \textit{The Best-Case (BC) Test:} The BC test is designed to give the predictor a best chance at making accurate predictions. It makes several strong assumptions. First, the one-feature-at-a-time assumption: in all the data given to the model (training and test), only a single parameter is being varied and that parameter is the only feature on which the model is trained. Second, the seen-configuration assumption: the model's training data always include datapoints from the scenario it is being asked to predict. Third, the no-contention assumption: to avoid variability due to contention, the workloads are run on dedicated EC2 instances. Fourth, the identical-inputs assumption: for a given input dataset size, the application's input data is identical across all experiments.
    
    \textit{The Beyond Best-Case (BBC) Test:} The BBC test systematically relax each of the assumptions imposed in the BC test, so as to study more realistic scenarios.
    
    \textit{The Best-of-Models Predictor:} In order to obtain a broad picture, a range of ML models are considered. For any given prediction test, MLSys computes the rMSRE for each ML model, and defines the best of models error (BoM-err) as the minimum rMSRE across all models considered.
    
    \textit{The Oracle Predictor:} In order to exclude the effect of poor ML model tuning, as well as obtaining a lower bound on the error rate we can expect from any ML model, an oracle predictor is used: it looks at all the data points in the test set that share the same feature values as the prediction task, and returns a prediction that will minimize the rMSRE for all these data points. If there is no variance at all in these data points, the Oracle error (O-err) will achieve zero.
    } \\
    
    \noindent {\textbf{Test Setup:} MLSys runs experiments against 13 real world applications, spanning data analytics, time series database, web services, etc.: Memcached, Nginx, Influxdb, Go-fasthttp, Spark (running Terasort, PageRank, logistic regression, KMeans, Word2Vec, FPGrowth, ALS), and Tensorflow. It aims to predict aggregate performance metric for each application, such as JCT and mean query latency.

    It selects six ML algorithms to train, including kNN, random forest regression, linear regression, linear SVM regression, kernelized SVM regression, and neural networks (MLP). Each ML model is tuned carefully.
    }
    
    \subsubsection{Tackling Irreducible Error}
    \label{sec:5.1.1}
    \begin{figure}[H]
    \centering
    \includegraphics[scale=0.55]{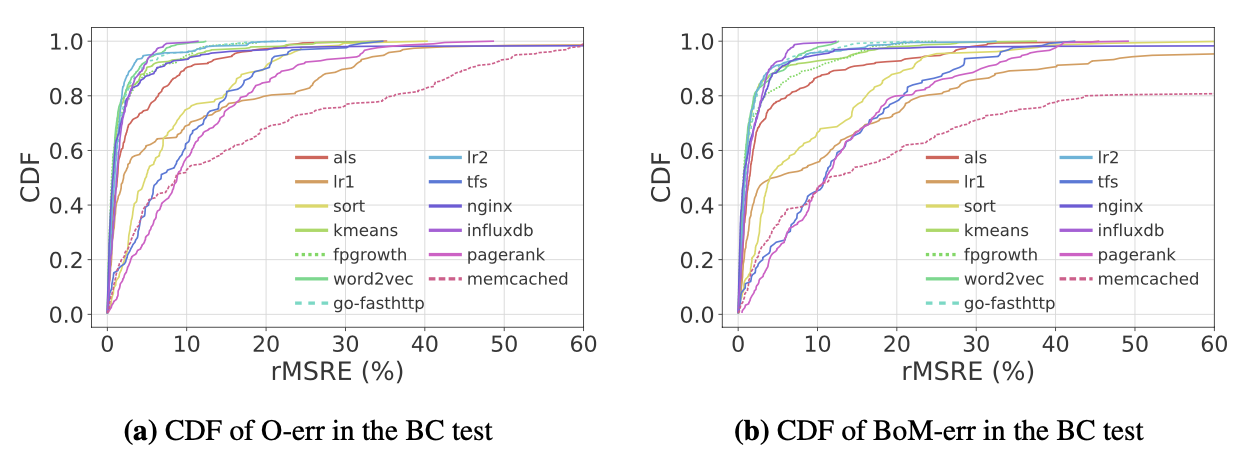}
    \caption{Results of Existing Applications}
    \label{fig:existing}
    \end{figure}  
    
    Given this methodology and set up, Figure~\ref{fig:existing} shows the results for existing applications. Since the BC test is extremely easy, we expected a very high accuracy: the O-err should be well under 5\% error. To our surprise, the results are on the contrary: in 5 of 13 applications, O-err is $>15\%$ for at least 20\% of prediction tasks. This high irreducible O-err suggests, the application's performance is so inherently variable that no predictor could predict performance with high accuracy, even on identical runs.
    
    To alleviate this problem, MLSys tried the first workaround: tackling irreducible error. It proceeds to find the root of such variability, which can be attributed to two common software design techniques: the use of randomization (e.g., load-balancing in TFS, task scheduling in Spark), and the use of system optimizations where a new mode of behavior is triggered by a threshold parameter (e.g., worker readiness in Spark, adaptive garbage collection in JVM). For the latter, we illustrate a more concrete example: by default, Spark launches an application once at least 80\% of its target worker nodes are ready, and the application partitions the input dataset based on the number of workers ready at this time. This optimization ensures resilience to failure and stragglers, but leads to variable parallelism and hence JCTs. Unfortunately, this variability cannot be captured by the static parameters/features we described above, since the exact degree of parallelism is affected by small differences in worker launch times, which is not known prior to runtime.
    
    \subsubsection{Results of Modified Applications}
    \label{sec:5.1.2}
    \begin{figure}[H]
    \centering
    \includegraphics[scale=0.55]{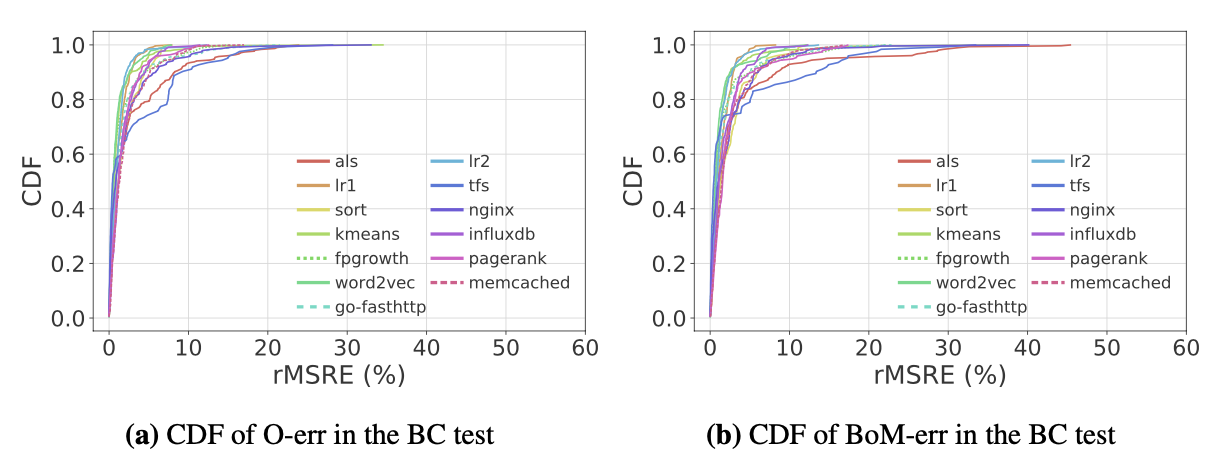}
    \caption{Results of Modified Applications}
    \label{fig:modified}
    \end{figure}  
    
    To justify that MLSys has found the root cause, we now present the results after removing these root causes. Moreover, if doing so promotes the prediction accuracy to a decent value, then we can imagine a workflow where the application developer identifies the root causes of irreducible errors and makes them configurable. However, this modification itself already concludes that \textit{ML is not a simple, i.e. easy to use, predictor for applications}.
    
    Figure~\ref{fig:modified} shows the results after removing the root causes. For BC test, all applications now have O-err well within 10\% for at least 90\% of their prediction tasks. In other words, 90-th percentile O-err is $<10\%$ for all 13 applications, and in fact, only two applications have O-err $>6\%$. Also, only two applications have 90-th percentile BoM-err $>6\%$. However, for BBC test (not shown here) where the seen-configuration-assumption is relaxed, multiple applications experience severe accuracy degradation: 10 out of the 13 applications have 90-th percentile BoM-err $> 10\%$; the 90-th percentile BoM-err of Memcached exceeds 60\%, while that of KMeans and TFS exceeds 25\%. This results concludes that \textit{ML is not a general predictor, i.e., suitable across a range of different applications, in real world settings}.
    
\begin{figure}[H]
    \centering
    \includegraphics[scale=0.4]{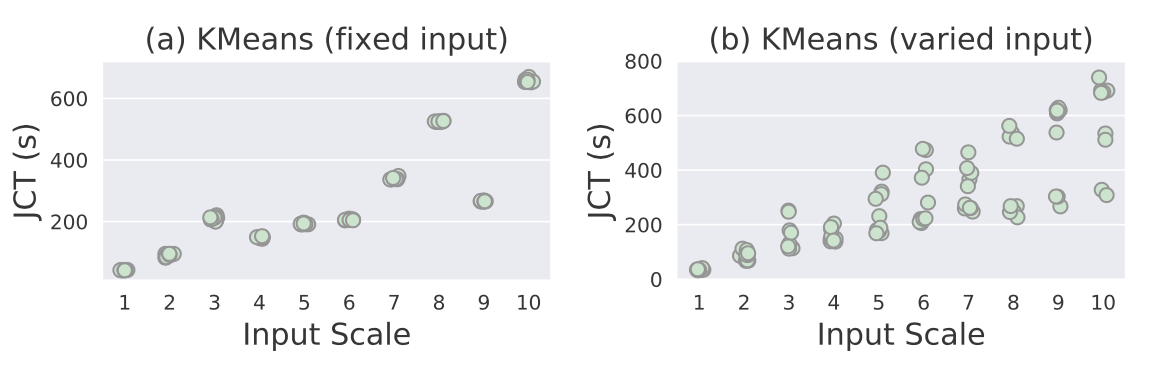}
    \caption{JCTs for KMeans}
    \label{fig:kmeans}
    \end{figure}

    Diving deeper on BBC prediction errors, MLSys further relaxed the identical-inputs assumption in addition to the seen-configuration-assumption and varied the input scale. It is shown that the applications most notably impacted are KMeans and logistic regression, since their performance is sensitive to the input data. For example, the number of iterations for KMeans depends on the actual values of the input data, resulting in a multi-modal behavior as is shown in Figure~\ref{fig:kmeans}. This entails high prediction errors for both fixed input and varied input.  
    
    \subsubsection{Probabilistic Predictions}
    Inspired by the multimodal performance distribution, MLSys tried the second workaround for performance variability: extending ML models to predict not just one performance value, but a probability distribution from which we derive $k$ possible values, with the goal that the true value is one of the $k$ predictions.

    MLSys extends neural network to Mixture Density Networks. It modifies the neural network to predict parameters for a Gaussian Mixture Model with $k$ components (mean and variance for each component and mixing coefficients). It uses negative log-likelihood of the data under the predicted GMM as the loss function to train the MDN. Similarly, random forest is extended to probabilistic random forests. Unfortunately, although this workaround helps reduce prediction errors, we continue to see scenarios with high error rates in the more realistic BBC tests. Thus, achieving a fully general predictor remains out of reach.
    
    \subsection{Evaluation}
    MLSys is a case study for Task 2. According to the taxonomy presented in Section~\ref{sec:taxonomy}, it adopts blackbox techniques for performance prediction. We summarize strengths and limitations of this field study as follows:\\
    
    \noindent \textbf{Strengths:}
    \begin{itemize}
        \item It is the first paper that \textit{comprehensively} evaluates whether ML-based prediction can \textit{simultaneously} offer high accuracy, easy-of-use and generality.
        \item It draws the conclusion that ML fails to serve as a simple and general predictor, while ML can be effective in several scenarios. This points a direction for system researchers: we must apply ML in a more \textit{nuanced} manner, by first identifying whether ML-based prediction is effective.

    \end{itemize}
    
    \noindent \textbf{Limitations:}
    \begin{itemize}
        \item Although it points out some root-causes, it does not provide a easy or systematic mechanism for operators to identify them. In addition, the elimination of variability hurts other goals such as resilience and efficiency that are crucial for distributed big data frameworks.
        \item This study does not provide the intuition on which ML model is best matched with which scenario. Instead, it simply brute-searches the models and the hyperparameters for each prediction task, which is cumbersome and not interpretable.
        \item The probabilistic prediction is only effective when the underlying trend is hard to learn due to the multi-modality. It does not work for other complex trends (e.g., in TFS and go-fasthttp).
    \end{itemize}
    
    \ifcsdef{mainfile}{}{\bibliography{../references/main}}

\pagebreak

\section{Making Disk Failure Predictions SMARTer}
\label{sec:smarter}
	
    In order to explore the second dimension of this proactive prediction, Lu et.al. conducted this field study (referred to as SMARTer) on disk failure prediction, which covers a total of 380,000 hard drives over a period of two months across 64 sites of a large datacenter operator. For the first time, SMARTer demonstrates that disk failure prediction can be made highly accurate by combing disk performance data and disk location data with disk monitoring data.
    
    \subsection{Overview}
    \label{sec:6.1}
    \textbf{Methodology:} In what follows, we first introduce the definition of failure, followed by the three types of data used throughout SMARTer. Given the complexity of disk failures, there is no common, agreed-upon universal definition of a disk failure. SMARTer considers a disk to be failed when there is a failed read/write operation and the disk cannot function properly upon restart. The failure label is tagged by the IT operators of the datacenter we study.
    
    SMART (Self-Monitoring, Analysis and Reporting Technology) attributes is the first source of data. It covers disk health measurements such as correctable errors, temperature, disk spin-up time, etc. The number of available SMART attributes is more than 50, but not all disks log all of the attributes at all times. The SMART attributes are collected and reported at per-day granularity.

    Performance is the second source of data. SMARTer collects two types of performance metrics maintained by the OS kernel: disk-level metrics and server-level metrics. Disk level performance metrics (12 in total) include IOQueue size, throughput, latency, the average waiting time for I/O operations, etc. Server level performance metrics (154 in total) include CPU activity, page in and out activities, etc. Performance metrics are reported at per-hour granularity, since hourly granularity is effective in improving the prediction quality.

    Location is the third source of data. Each disk has four levels of location markers associated with it: site, room, rack, and server.

    To evaluate the effectiveness of different features---SMART attributes (S), Performance metrics (P), and location markers (L), six different feature groups are considered: SPL, SL, SP, PL, S, and P. Precision, recall, F-measure, and Matthews correlation coefficient are used as metrics for evaluation.
    
    \subsubsection{Selection of Attributes}
    \label{sec:6.1.1}
    Since the storage overhead of all performance metrics can become significant at scale and over time, SMARTer leverages J-Index to down-select features. After features are normalized to the scale of 0-1, SMARTer sets a series of threshold candidates for each feature with a step of 0.01, starting from 0 until 1, and calculates the value of corresponding J-Index: since the distribution of failed disk and healthy disk along the feature is known, and the threshold is also known, we can calculate the TP/TN/FP/FN (predict the disk failure if only according to that feature and threshold) and thus J-Index. A higher J-Index means the corresponding threshold candidate is more distinguishable to identify failure disks from healthy disks. Then, the highest J-Index across thresholds is selected for every feature, and features with higher J-Index are selected as input for the ML models. Eventually, 14 SMART attributes, 12 disk level performance metrics, 18 server level performance metrics are selected.
    
    \subsubsection{ML Models}
    \label{sec:6.1.2}
    The input to a ML model is a time sequence of the selected attributes, i.e. multiple and fixed length readings. The output of a ML model is a binary label, indicating whether this disk will fail in the next ten days.
    
\begin{figure}[H]
    \centering
    \includegraphics[scale=0.35]{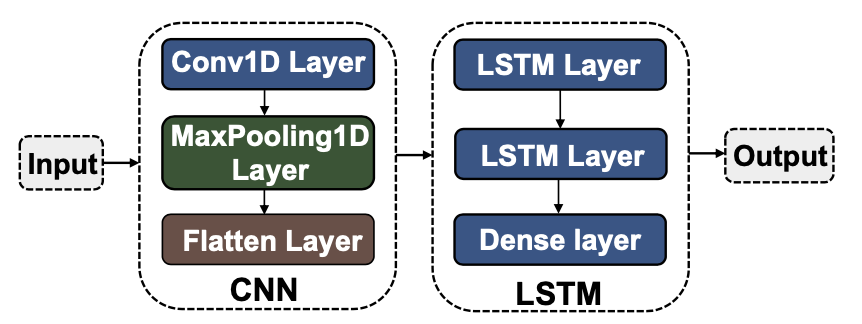}
    \caption{Architecture of CNN-LSTM}
    \label{fig:cnnlstm}
    \end{figure}

    SMARTer implemented and tuned five ML models, including Bayes, RF, GBDT, LSTM, CNN-LSTM. Among them, CNN-LSTM is a newly proposed model (as shown in Figure~\ref{fig:cnnlstm}), where CNN offers advantages in selecting better features, while LSTM is effective at learning sequential data.
    
    \subsubsection{Findings}
    \label{sec:6.1.3}
    Below we summarize the important findings and lessons from SMARTer:\\
    
\begin{figure}[H]
    \centering
    \includegraphics[scale=0.5]{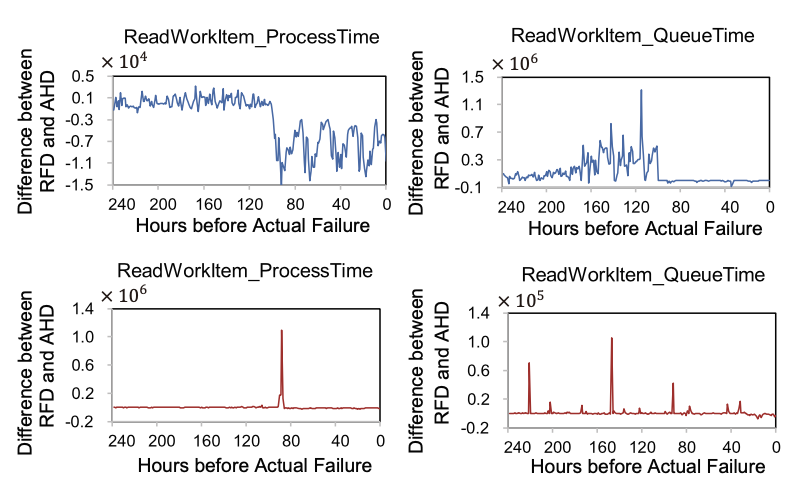}
    \caption{Pattern in Performance Differences}
    \label{fig:error perf}
    \end{figure}

    \noindent \textbf{Feature Importance:} The SPL feature group performs the best across all ML models, showing that performance and location features are critical for improving the effectiveness of disk failure prediction. More specifically, SMART attributes do not always have the strong predictive capability of making predictions at longer prediction horizon (e.g., 10 days ahead), because the changes in values of SMART attributes are often noticeable only a few hours before the actual failure. On the other hand, the performance may exhibit more variations long before the actual failure. Figure~\ref{fig:error perf} shows the performance difference between raw failed disk and averaged healthy disks. In Figure~\ref{fig:error perf}, some failed disks have a similar value to healthy disks at first, but then their behavior becomes unstable when approaching the impending failure; some other failed disks report a sharp impulse before they fail, as opposed to a linger erratic behavior. Interestingly, the effect of location information is pronounced only in the presence of performance features.
    
    \noindent \textbf{Model Selection:} While there is no single model winner across different feature groups, CNN-LSTM performs close to the best in all the situations. However, when performance features are not available, traditional tree-based models (RF and GBDT) can perform roughly as well as complex neural networks. Considering the large training overhead of neural networks, RF and GBDT are more preferred in this case.\\
    
    \noindent \textbf{Portability:} If we simply try to train the model on one datacenter site and port it to another datacenter site (i.e., test on another unseen site), the MCC score can drop significantly. However, if a model is trained on multiple datacenter sites before testing on a new unseen one, it provides high prediction accuracy, especially when using CNN-LSTM.
    
    \subsection{Evaluation}
    SMARTer is a case study for Task 3. According to the taxonomy presented in Section~\ref{sec:taxonomy}, it is an offline scheme for failure prediction. We summarize strengths and limitations of this field study as follows:\\
    
    \noindent \textbf{Strengths:}
    \begin{itemize}
        \item It is the first paper that demonstrates the strong predictive capability of SMART attributes combined performance metrics and location markers. This encourages researchers to pay more attention to the source of data instead of focusing on ML algorithms alone.
        \item It provides several actionable insights and trade-off lessons learned from a large scale, real world study.
    \end{itemize}
    
    \noindent \textbf{Limitations:}
    \begin{itemize}
        \item It does not consider the heterogeneity of hardware in deployment, i.e., different vendors and different drive types. Every component in the ML pipeline, i.e., feature engineering, model selection, model training and evaluation, is agnostic to the underlying device type.
        \item It directly chooses a simple J-Index based feature selection algorithm without even considering other fancier ones, such as random forest, XGBoost, Pearson correlation, etc. 
    \end{itemize}
    
    \ifcsdef{mainfile}{}{\bibliography{../references/main}}

\pagebreak

\section{Discussion}
\label{sec:discussion}
	
    To conclude our survey paper, we discuss the remaining challenges along with possible future directions for proactive predictions in storage systems.
    
    \begin{itemize}
        \item \textbf{Masking ML Failure in Systems:} As machine learning can never achieve 100\% accuracy, how should "ML-for-system" solutions mask the cases that cause big disruption of the system and machine learning fails to catch? LinnOS and SMARTer has made a successful attempt in marrying ML and existing system solutions (e.g., hedged request, redundancy scheme), but more attention is deserved.
        \item \textbf{AutoML:} If we apply ML to a general setting across a range of applications, just like what the MLSys paper did, a natural question is what model and hyperparameters should we choose for each different setting. Currently, MLSys simply conducts brute search, which is cumbersome and not scalable. AutoML techniques are likely to help in selecting the best model for different applications, drive types, etc. 
        \item \textbf{Model Portability:} Due to the heterogeneity of hardware and the frequent expansion of datacenters, portable ML models would be highly beneficial. The SMARTer paper showed that we can train the model on one site and deploy it on another with reasonable accuracy. However, can we train the model on certain device type and deploy it on another, e.g., training on SSDs and deploying on HDDs? How to develop more portable models? These questions are to be answered.
        \item \textbf{Lifelong Learning:} Due to the everlasting changes in observed workload of each device, long term deployment of ML models in systems requires them to learn from new observations continuously. While LinnOS utilizes model recalibration to adapt to the significant changes, which actually did not happen in LinnOS's experiment, more interesting questions need to be answered. For example, when we encounter an unseen workload, can we pull a model out from a repository of models which are already trained for different workloads? Can we utilize online ML algorithms to adapt to the changes? How to not forget what is already learnt?
        \item \textbf{Unbalanced Data:} The problem of data imbalance is particularly severe in failure prediction for storage systems, since drive failures (positive samples) only account for less than 10\% of the data. Thus, high false negative rates are common for failure prediction. The SMARTer paper also observed this with the S group. However, the false negative rate miraculously decreased for the SPL group without further model modification. Several questions are still not explored in the systems field: Does down-sampling negative samples help boost prediction accuracy? What ML techniques can we utilize to cope with data imbalance?
    \end{itemize}
    
    \ifcsdef{mainfile}{}{\bibliography{../references/main}}

\begin{thebibliography}{10}

\bibitem{hao2016tail}
Mingzhe Hao, Gokul Soundararajan, Deepak Kenchammana-Hosekote, Andrew~A Chien,
  and Haryadi~S Gunawi.
\newblock The tail at store: A revelation from millions of hours of disk and
  {SSD} deployments.
\newblock In {\em 14th USENIX Conference on File and Storage Technologies (FAST
  16)}, pages 263--276, 2016.

\bibitem{lu2020making}
Sidi Lu, Bing Luo, Tirthak Patel, Yongtao Yao, Devesh Tiwari, and Weisong Shi.
\newblock Making disk failure predictions smarter!
\newblock In {\em 18th USENIX Conference on File and Storage Technologies (FAST
  20)}, pages 151--167, 2020.

\bibitem{han2021depth}
Shujie Han, Patrick~PC Lee, Fan Xu, Yi~Liu, Cheng He, and Jiongzhou Liu.
\newblock An in-depth study of correlated failures in production {SSD}-based
  data centers.
\newblock In {\em 19th USENIX Conference on File and Storage Technologies (FAST
  21)}, pages 417--429, 2021.

\bibitem{xugeneral}
Fan Xu, Shujie Han, Patrick~PC Lee, Yi~Liu, Cheng He, and Jiongzhou Liu.
\newblock General feature selection for failure prediction in large-scale {SSD}
  deployment.

\bibitem{dean2013tail}
Jeffrey Dean and Luiz~Andr{\'e} Barroso.
\newblock The tail at scale.
\newblock {\em Communications of the ACM}, 56(2):74--80, 2013.

\bibitem{hao2017mittos}
Mingzhe Hao, Huaicheng Li, Michael~Hao Tong, Chrisma Pakha, Riza~O Suminto,
  Cesar~A Stuardo, Andrew~A Chien, and Haryadi~S Gunawi.
\newblock Mittos: Supporting millisecond tail tolerance with fast rejecting
  {SLO}-aware {OS} interface.
\newblock In {\em Proceedings of the 26th Symposium on Operating Systems
  Principles}, pages 168--183, 2017.

\bibitem{hao2020linnos}
Mingzhe Hao, Levent Toksoz, Nanqinqin Li, Edward~Edberg Halim, Henry Hoffmann,
  and Haryadi~S Gunawi.
\newblock Linnos: Predictability on unpredictable flash storage with a light
  neural network.
\newblock In {\em 14th USENIX Symposium on Operating Systems Design and
  Implementation (OSDI 20)}, pages 173--190, 2020.

\bibitem{mahdisoltani2017proactive}
Farzaneh Mahdisoltani, Ioan Stefanovici, and Bianca Schroeder.
\newblock Proactive error prediction to improve storage system reliability.
\newblock In {\em 2017 USENIX Annual Technical Conference (USENIX ATC 17)},
  pages 391--402, 2017.

\bibitem{han2020toward}
Shujie Han, Patrick~PC Lee, Zhirong Shen, Cheng He, Yi~Liu, and Tao Huang.
\newblock Toward adaptive disk failure prediction via stream mining.
\newblock In {\em Proceedings of IEEE ICDCS}, 2020.

\bibitem{barroso2018datacenter}
Luiz~Andr{\'e} Barroso, Urs H{\"o}lzle, and Parthasarathy Ranganathan.
\newblock The datacenter as a computer: Designing warehouse-scale machines.
\newblock {\em Synthesis Lectures on Computer Architecture}, 13(3):i--189,
  2018.

\bibitem{zaharia2012resilient}
Matei Zaharia, Mosharaf Chowdhury, Tathagata Das, Ankur Dave, Justin Ma, Murphy
  McCauly, Michael~J Franklin, Scott Shenker, and Ion Stoica.
\newblock Resilient distributed datasets: A fault-tolerant abstraction for
  in-memory cluster computing.
\newblock In {\em 9th USENIX Symposium on Networked Systems Design and
  Implementation (NSDI 12)}, pages 15--28, 2012.

\bibitem{zaharia2010spark}
Matei Zaharia, Mosharaf Chowdhury, Michael~J Franklin, Scott Shenker, Ion
  Stoica, et~al.
\newblock Spark: Cluster computing with working sets.
\newblock {\em HotCloud}, 10(10-10):95, 2010.

\bibitem{wang2004storage}
Mengzhi Wang, Kinman Au, Anastassia Ailamaki, Anthony Brockwell, Christos
  Faloutsos, and Gregory~R Ganger.
\newblock Storage device performance prediction with {CART} models.
\newblock In {\em The IEEE Computer Society's 12th Annual International
  Symposium on Modeling, Analysis, and Simulation of Computer and
  Telecommunications Systems, 2004.(MASCOTS 2004). Proceedings.}, pages
  588--595. IEEE, 2004.

\bibitem{yin2006empirical}
Li~Yin, Sandeep Uttamchandani, and Randy Katz.
\newblock An empirical exploration of black-box performance models for storage
  systems.
\newblock In {\em 14th IEEE International Symposium on Modeling, Analysis, and
  Simulation}, pages 433--440. IEEE, 2006.

\bibitem{varki2004issues}
Elizabeth Varki, Arif Merchant, Jianzhang Xu, and Xiaozhou Qiu.
\newblock Issues and challenges in the performance analysis of real disk
  arrays.
\newblock {\em IEEE Transactions on Parallel and Distributed Systems},
  15(6):559--574, 2004.

\bibitem{anderson2002hippodrome}
Eric Anderson, Michael Hobbs, Kimberly Keeton, Susan Spence, Mustafa Uysal, and
  Alistair~C Veitch.
\newblock Hippodrome: Running circles around storage administration.
\newblock In {\em FAST}, volume~2, pages 175--188, 2002.

\bibitem{hsu2016inside}
Chin-Jung Hsu, Rajesh~K Panta, Moo-Ryong Ra, and Vincent~W Freeh.
\newblock Inside-out: Reliable performance prediction for distributed storage
  systems in the cloud.
\newblock In {\em 2016 IEEE 35th Symposium on Reliable Distributed Systems
  (SRDS)}, pages 127--136. IEEE, 2016.

\bibitem{fu2021use}
Silvery Fu, Saurabh Gupta, Radhika Mittal, and Sylvia Ratnasamy.
\newblock On the use of ml for blackbox system performance prediction.
\newblock In {\em NSDI}, pages 763--784, 2021.

\bibitem{ousterhout2017monotasks}
Kay Ousterhout, Christopher Canel, Sylvia Ratnasamy, and Scott Shenker.
\newblock Monotasks: Architecting for performance clarity in data analytics
  frameworks.
\newblock In {\em Proceedings of the 26th Symposium on Operating Systems
  Principles}, pages 184--200, 2017.

\bibitem{venkataraman2016ernest}
Shivaram Venkataraman, Zongheng Yang, Michael Franklin, Benjamin Recht, and Ion
  Stoica.
\newblock Ernest: Efficient performance prediction for large-scale advanced
  analytics.
\newblock In {\em 13th USENIX Symposium on Networked Systems Design and
  Implementation (NSDI 16)}, pages 363--378, 2016.

\bibitem{yadwadkar2014wrangler}
Neeraja~J Yadwadkar, Ganesh Ananthanarayanan, and Randy Katz.
\newblock Wrangler: Predictable and faster jobs using fewer resources.
\newblock In {\em Proceedings of the ACM Symposium on Cloud Computing}, pages
  1--14, 2014.

\bibitem{verma2011aria}
Abhishek Verma, Ludmila Cherkasova, and Roy~H Campbell.
\newblock Aria: automatic resource inference and allocation for mapreduce
  environments.
\newblock In {\em Proceedings of the 8th ACM international conference on
  Autonomic computing}, pages 235--244, 2011.

\bibitem{xiao2018disk}
Jiang Xiao, Zhuang Xiong, Song Wu, Yusheng Yi, Hai Jin, and Kan Hu.
\newblock Disk failure prediction in data centers via online learning.
\newblock In {\em Proceedings of the 47th International Conference on Parallel
  Processing}, pages 1--10, 2018.

\bibitem{xu2018improving}
Yong Xu, Kaixin Sui, Randolph Yao, Hongyu Zhang, Qingwei Lin, Yingnong Dang,
  Peng Li, Keceng Jiang, Wenchi Zhang, Jian-Guang Lou, et~al.
\newblock Improving service availability of cloud systems by predicting disk
  error.
\newblock In {\em 2018 USENIX Annual Technical Conference (USENIX ATC 18)},
  pages 481--494, 2018.

\bibitem{alter2019ssd}
Jacob Alter, Ji~Xue, Alma Dimnaku, and Evgenia Smirni.
\newblock {SSD} failures in the field: symptoms, causes, and prediction models.
\newblock In {\em Proceedings of the International Conference for High
  Performance Computing, Networking, Storage and Analysis}, pages 1--14, 2019.

\end{thebibliography}


\end{document}